\pdfoutput=1
\documentclass[aps,prl,preprint,groupedaddress]{revtex4-1}

 \usepackage{graphicx}
\begin{document}


\title{Experimental investigation of the trachea oscillation and its role in the pitch production.}

\author{G. Buccheri}
\affiliation{Scuola di dottorato in Scienze e Tecnologie, dell'Informazione dei Sistemi Complessi e dell'Ambiente, Universit\`a degli Studi di Salerno, Via Giovanni Paolo II, I-84084 Fisciano (SA), Italy}

\author{E. De Lauro}
\author{S. De Martino}\thanks{Corresponding Author. Electronic Address:demartino@sa.infn.it}
\author{M. Falanga}
\affiliation{Dipartimento di Ingegneria dell'Informazione, Ingegneria Elettrica e Matematica Applicata-DIEM, Universit\`a degli Studi di Salerno, Via Giovanni Paolo II, I-84084 Fisciano (SA), Italy}

\date{\today}

\begin{abstract}
Several experiments have been performed to investigate the mechanical vibrations associated with
trachea and larynx when Italian vowels are emitted. The mechanical measurements have been made by using two laser Doppler vibrometers (based on the well-known not-invasive optical measurement technique) coupled with the acoustic field measured by high-quality certified microphones. The recorded signals are analyzed by using well-established methods in time and frequency domains. The signals of the mechanical vibrations along the trachea and the larynx are compared with those of the acoustic ones. Focusing the attention of the signals' onsets, we can observe an upward   propagation of the mechanical vibrations for which it is possible to estimate a delay between the traces. We observe that the mechanical oscillations at the trachea start before the larynx and acoustic oscillations. Moreover these tracheal oscillations are self-oscillations in time and are associated with the pitch production, indicating a further hydrodynamic instability at trachea. This leads to new insights in the mechanism controlling the pitch in the speech.
\end{abstract}


\maketitle


\section{Introduction}

The understanding of the mechanism by which the brain organizes the complex temporal activity of speech is relevant both for speech and brain. But in addition to the neural control, a further level of temporal organization is provided by oscillatory dynamics, which is intrinsic to the vocal organ. Looking at these intrinsic oscillations, from the very beginning, physical models of voice production have been developed within a nonlinear framework. Indeed, observing the behavior of larynx within the complex system of vocal tract (lung-trachea, larynx, upper vocal tract), at the beginning of nineteenth century, Airy \cite{Airy 1826} introduced his well-known delay-equation, today used as a simple epidemic model. Airy's work provided the theoretical framework associated to a series of experimental papers of Willis \cite{Willis 1826} regarding larynx, organ pipes and their similarity.
The next step in this direction must wait the contributions of Wegel about a century later \cite{Wegel 1930}. He introduced a suitable Lagrangian, whose variation provides an equations' system, whose solutions are self-oscillations. Thus, the main idea is that the voice is produced as a result of dynamical instabilities generated by larynx. So the model of larynx has received many improvements along the time, thanks to many relevant contributions (see among the others, \cite{Fant 1970,Flanagan 1972,Tizte 1973,Tizte 1974,Kent1993}). The standard model considers the larynx as a source and the upper vocal tract as a filter (source-filter model) but Teager and Teager \cite{Teager1983,Teager1990} have shown that within the mouth the flows cannot be traced to laminar but they develop eddies. The subsequent efforts have been dedicated to refine larynx model and to introduce a suitable coupling with mouth \cite{Lucero2006} as well.
However, none of these attempts can be considered fully successful, and we cannot claim to have a good synthetic voice even for a simple vowel, namely the models produced in a speech context have an immediate experimental reply from synthesizing. Another relevant aspect is associated to the detection of the pitch. Its role is considered relevant in characterizing people but not clearly understood especially regarding its generation.

In this work, we show the results of a series of experiments and measures looking at thorax mechanical vibrations. The respiratory system has been thought of infrequently as an acoustic milieu but many experiments, with acoustical measurements apparatus, have been performed along the time \cite{VandeBerg1958} and an electrical analogue of trachea-lung system has been produced. So, it can be interesting to revise this topic with innovative measurements, i.e,. by using laser Doppler vibrometer and also performing new analysis. It can have relevance in nonlinear systems to introduce a not constant flow as a source. We remember that the air flow coming out from the lung has been considered to produce a slowly varying tracheal air pressure that at first approximation does not influence the voice production. But clinical experience suggests that tracheal pull generating an abductive glottal force can be relevant (see, e.g., \cite{Iwarsson1998} and references therein). We concentrate our attention to vowels, which can be considered the simplest example of speech production.

\section{Dataset and feature extraction}
All the experiments have been performed in an anechoic chamber considering men and women. We have simultaneously measured the mechanical oscillations (velocity) in two variable thorax points by means of two laser Doppler vibrometers displaced of 0.11m (see Fig. 1) and the acoustic emission with a set of microphones (here we report only the trace relative to channel CH3 for convenience). The experiments have been conducted in a variety of conditions, namely with different male and female individuals and in different situations: i.e. voice normalized or not, whispering and silent. In this first paper, we describe very general results independent of the particular experimental condition and regarding voiced vowels.

\begin{figure}[!]
\label{Apparatus}
\includegraphics[width=12cm]{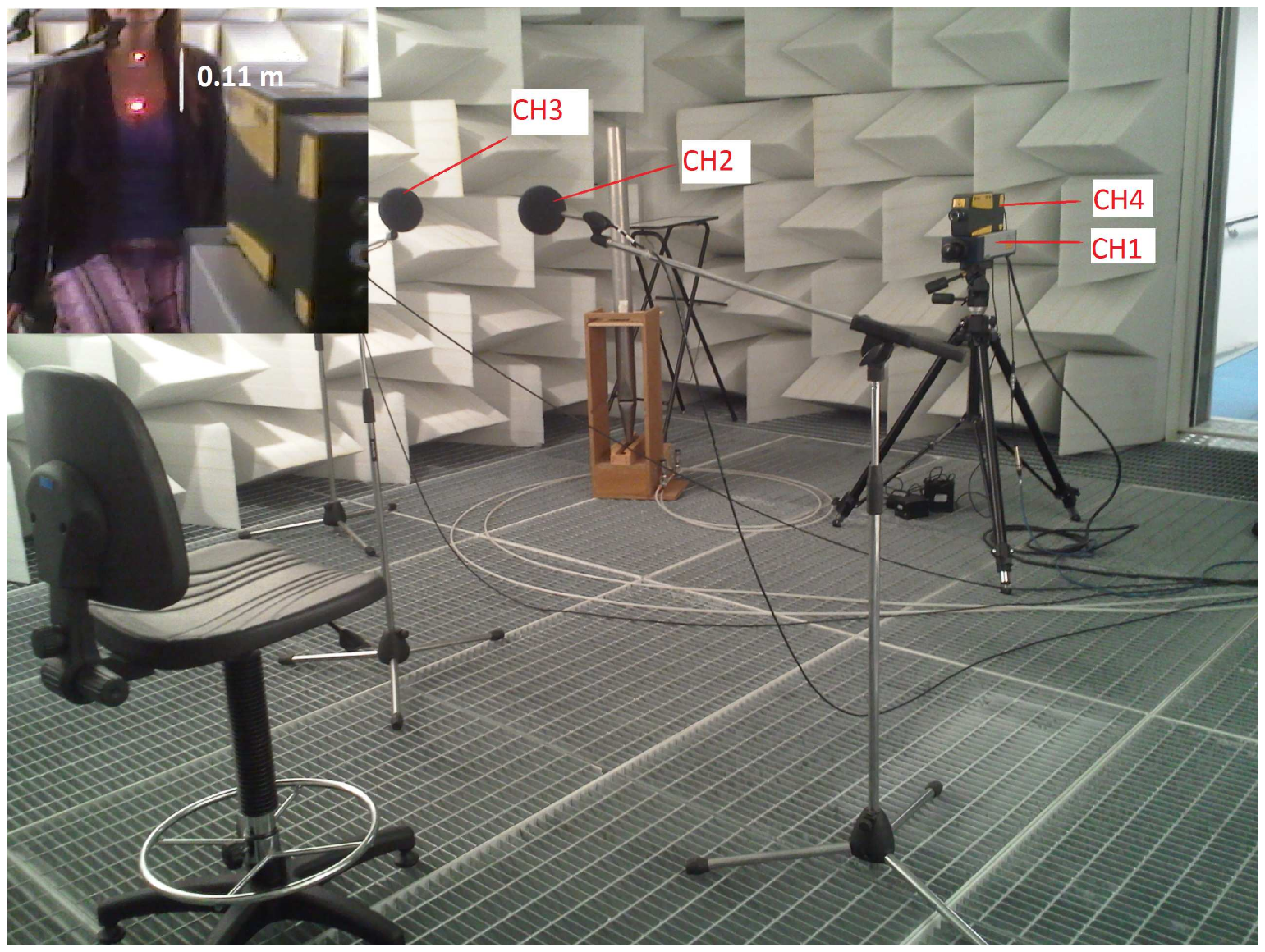}
\caption{Configuration of the experimental apparatus. The mechanic oscillations are measured simultaneously in two points of the thorax by using two laser Doppler vibrometers, (Polytec OFV-5000 and PDV-100) with frequency band acquisition up to 20KHz, full scale $\pm$4V, and sensitivity $5\frac{\frac{mm}{s}}{V}$. Moving the laser of 0.01 mm we scan the entire trachea. The acoustic field has been recorded simultaneously by using two microphones.}
\end{figure}

In Fig. 2, we report as an example acoustic oscillations (CH3) and mechanical ones (CH1, CH4) relative to a female respectively for five Italian voiced vowels in the sequel [a], [e], [i], [o], [u], together with the spectrogram in which the formants are easily recognizable. The mechanical recorded signals are very significant and if one listens these signals comparing them with the acoustical ones, it is easy to appreciate a trace of recognition of the specific speaker. For the acoustic part, formants are very well shown and in agreement with the spectrogram reported in \cite{Ferrero1978,Ferrero1996}, whereas in the laser signals for all the vowels only a low frequency is present with a period corresponding to the pitch. Indeed, superimposing the recorded mechanical signal (CH1) to the correspondent acoustical one (CH3), we observe a seeming prosodic modulation of the mechanical oscillation with respect to the acoustic one.
Additional information on the pitch can be derived from methods based on the correlation. Considering that the signals are associated to nonlinear systems, we adopt the Average Mutual Information (AMI) \cite{Fraser1986} as estimator of the pitch extraction. Specifically, from the comparison between the microphone and laser signals, we have a very clear indication of the pitch and of all the other contributions in the produced vowel. In Fig. 3b, we plot both the AMIs: the second maximum peak gives information on the pitch and it can be seen how simple is the shape of the laser compared with the microphone.

\begin{figure}
\label{spettrogramma}
\includegraphics[width=12cm]{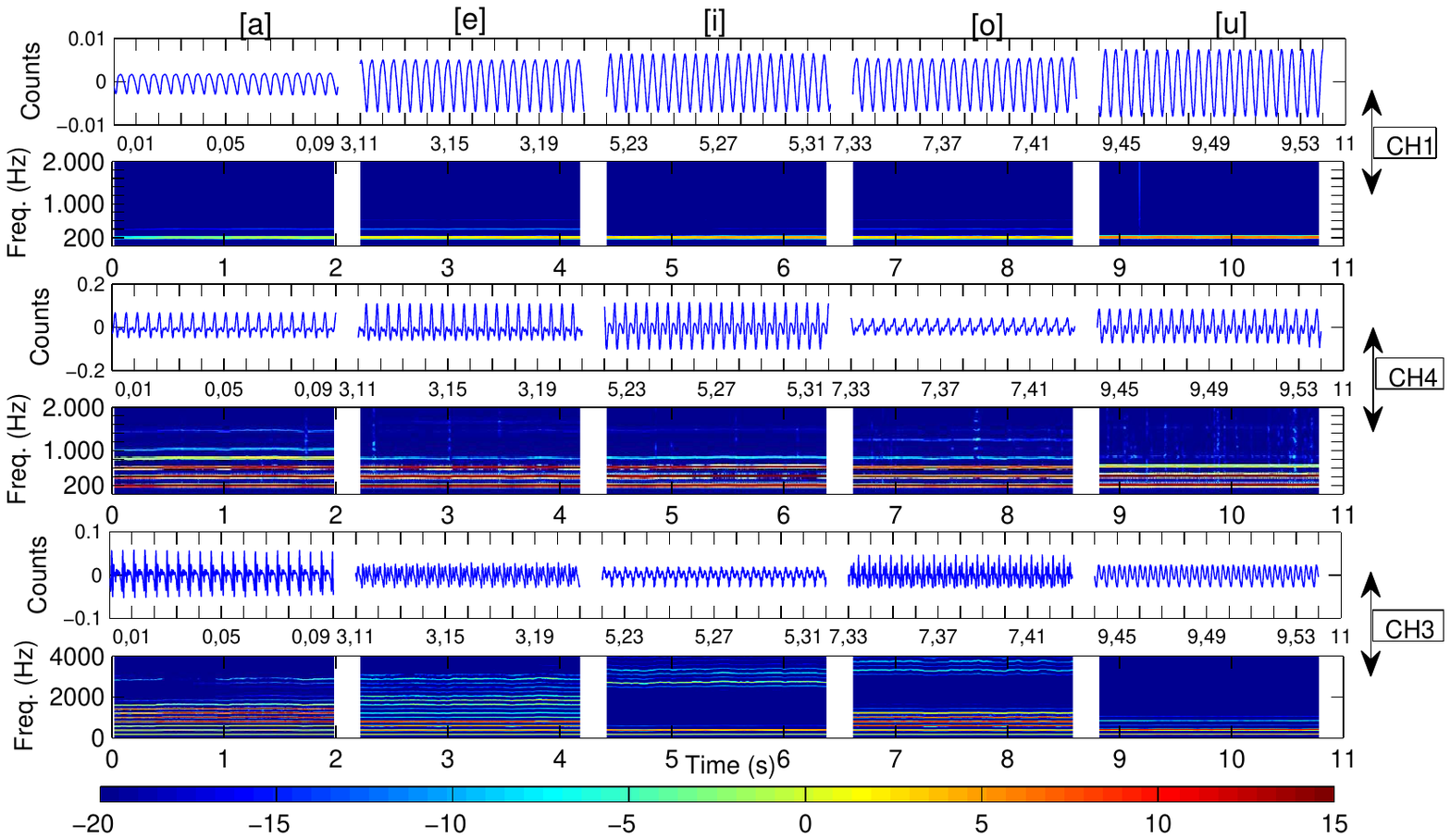}
\caption{a) Waveforms of Italian vowels regarding a female 35 years old acquired by two lasers (CH1, CH4) and a microphone (CH3) with relative spectrogram b).}
\end{figure}

\begin{figure}
\label{mutual}
\includegraphics[width=12cm]{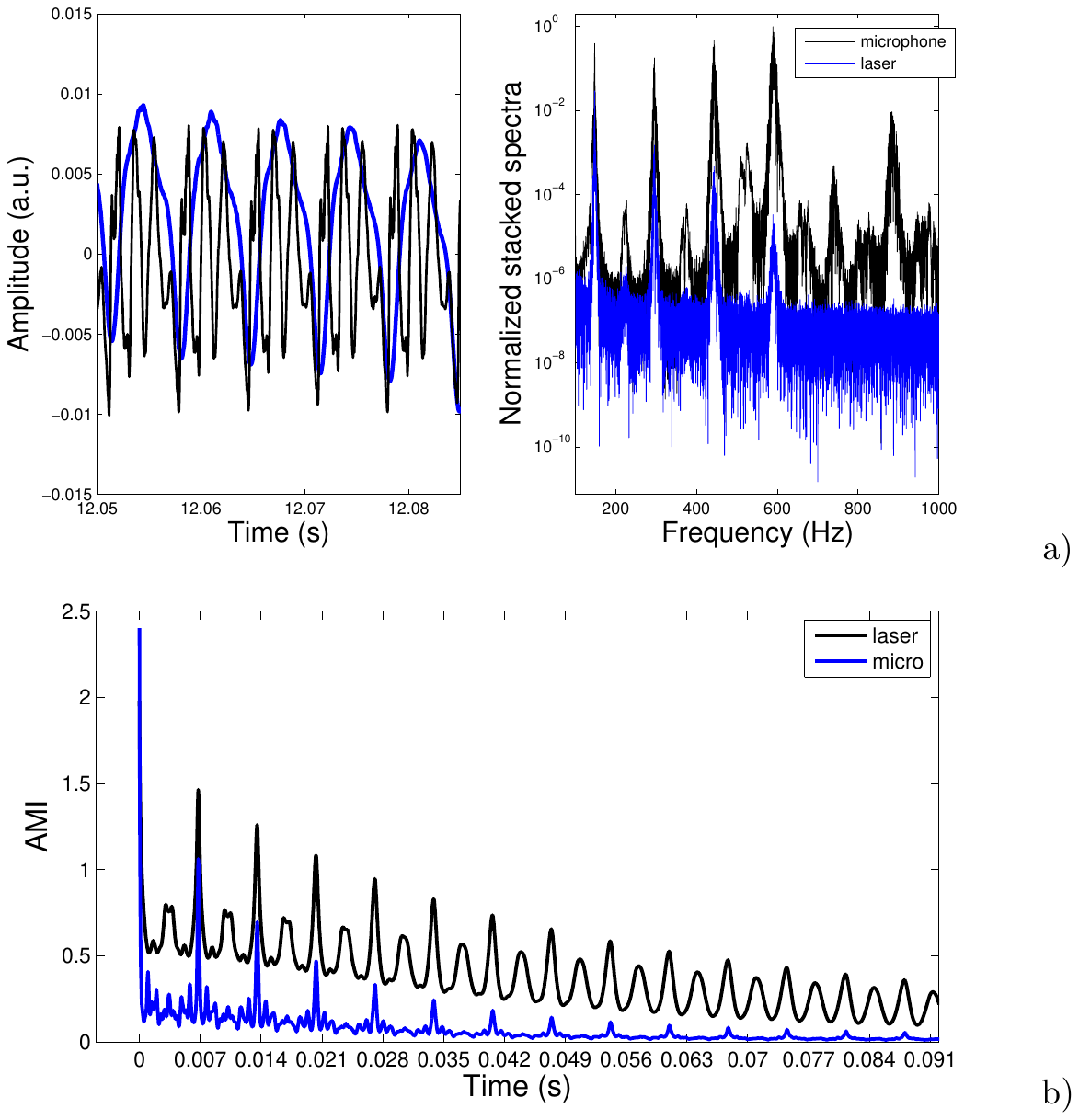}
\caption{a)Superposition of the microphone (CH3) and laser (CH1) signal in normalized units relative to [e] vowel emitted by a male and the relative power spectral density. As it can be seen, CH1 modulates CH3 and a delay between the traces can be revealed. This feature will be deeply investigated in the following. The same occurs for female if one compares the acoustic and mechanical signals in Fig. 4; b) AMI evaluated both for acoustic and mechanical vibrations: the second maximum indicates the pitch and coincides in both cases.}
\end{figure}

As usual, the pitch is extracted in time domain as distance between two consecutive maxima of the acoustic signal. This inter-time analysis is applied to both microphones and laser signals with the aim to recover the statistics, i.e. the average value (that usually correspond to the pitch) and the shape of the distribution. In Fig. 4, we can see the pitch extracted from both microphone (CH3) and laser signals (CH1, CH4). It is noteworthy to observe that there is a difference, only within the error range, relatively to the values obtained by looking at acoustic (as usual) or mechanical vibrations (now introduced). Moreover, we can see the distribution of the inter-times are more peaked for lasers with respect to the acoustic one, meaning that it is a basic signal. Indeed, the Gaussianity (sub/super) of the distribution can be measured by means of a parameter, the variability coefficient $C_{V}$, defined as follows:

$$C_{V}=\frac{\sigma_{\Delta t}}{\overline{\Delta t}},$$ where $\sigma_{\Delta t}$ is the standard deviation and $\overline{\Delta t}$ is the mean value of the inter-times. $C_{V}=1$ is for a Poissonian process, whereas $C_{V}>1$ is for a clustered process and $C_{V}=0$ is for a periodic one. The limit $C_{V} \rightarrow \infty$ indicates uniform distributions.
The three $C_{V}s$ are 0.02, 0.4, 0.1 for CH1, CH3 and CH4 respectively, with CH1 that is one order of magnitude smaller than the others. This indicates a major complexity at glottis signal with respect to the signal in the trachea, which can be rightly considered as basic. Moreover, the departure of the value from zero strengthens the indication that the dynamics is a self-oscillation.

\begin{figure}
\label{innter_picth_woman}
\includegraphics[width=12cm]{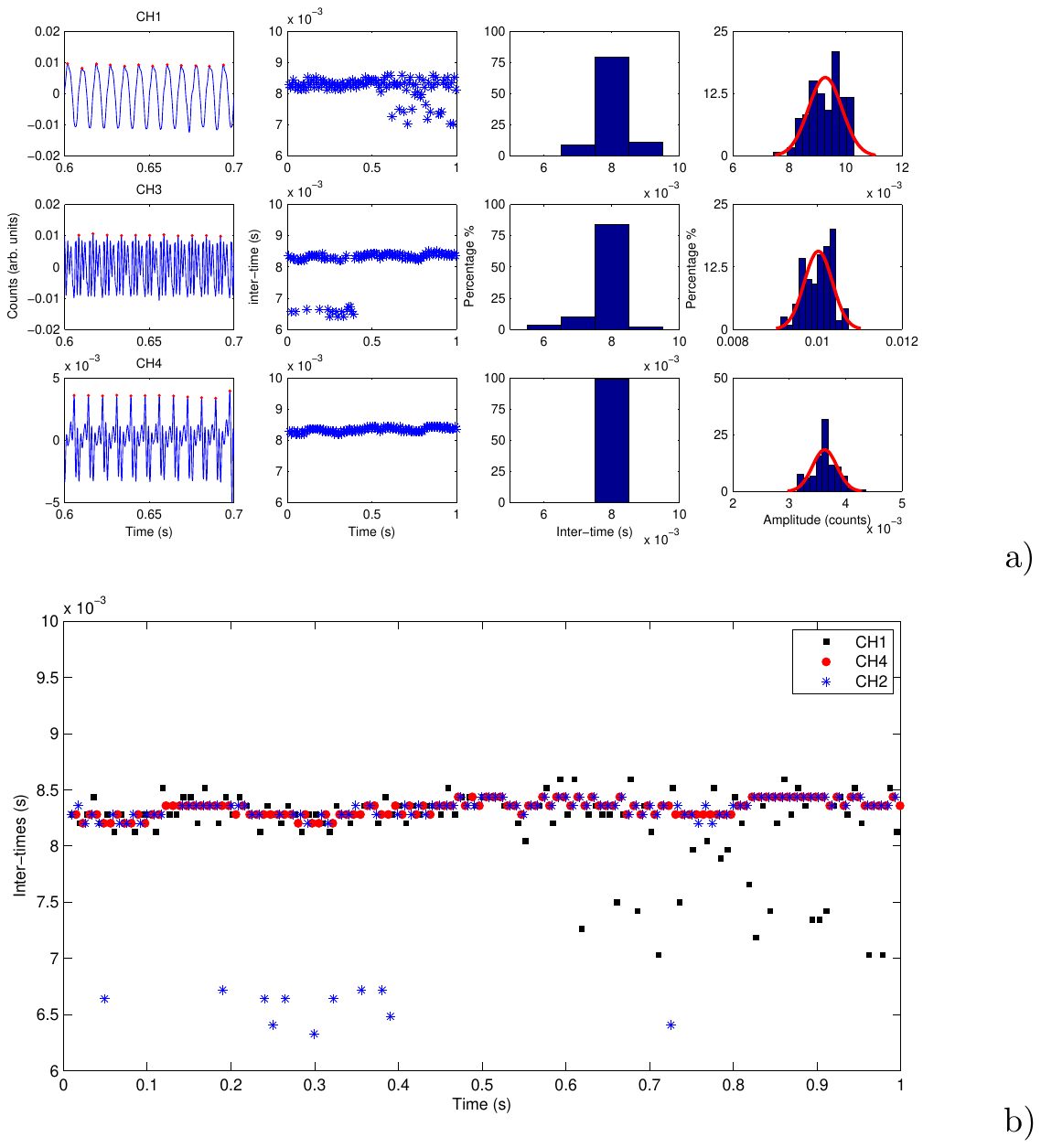}
\caption{Intertime analysis: extraction of pitch for [e] vowel for a woman.}
\end{figure}

\newpage
A question immediately arises: Are these mechanical signals an effect of pressure gradient at the glottis?
In order to reveal the origin (source) of such a signal, it is necessary to deep investigate the onsets of laser signals at CH1 and CH4. Indeed, simultaneous measures with a high sampling rate (102.4 kHz our choice) allow to study transient signals and to trace the propagation thus inferring on the source. As a result, Fig. 5 clearly shows a time delay between the signals recorded at the two selected points reveling an upwards flow (from CH1 to CH4). In details, we report the extraction of the onset of the mechanical oscillations both for a male and a female in emitting [e] vowel. We estimate the time delay between the two measures by means of the cross-correlation analysis.
\begin{figure}
\label{enza_attack}
\includegraphics[width=12cm]{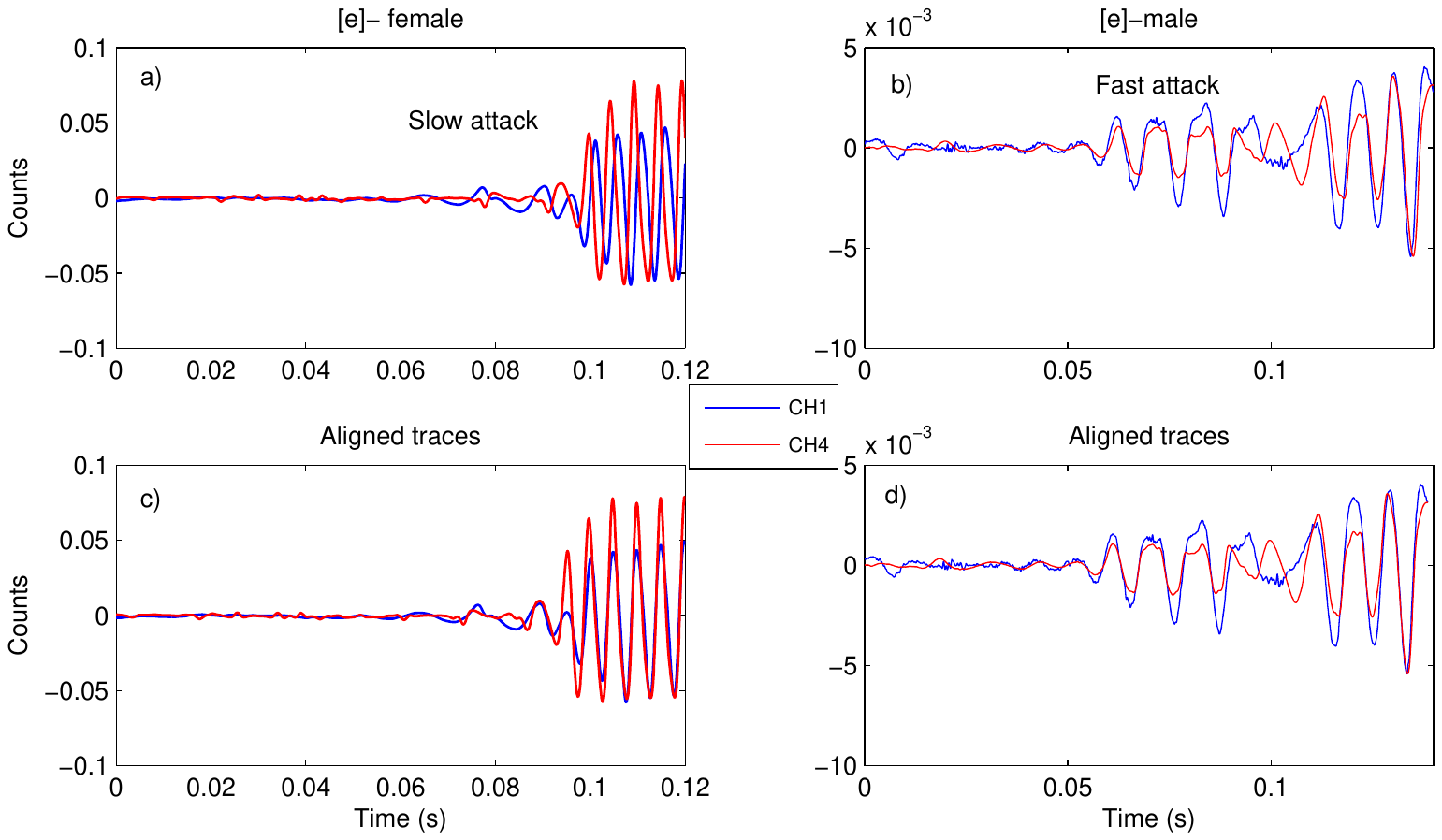}\\
\caption{Onset of mechanical oscillations: a) for a female and c) aligned traces by means of cross-correlation analysis; b) for a man and d)aligned traces by means of cross-correlation analysis. The time lag equal to 0.035s from cross-correlation for a female is one order higher than the one for a man, producing a velocity of about 30m/s. For a man the time lag estimated by cross-correlation is 4*10-4s and the velocity is about 270m/s. This situation remind us the behaviour of the attack in a playing organ pipe. In the first case we can speak of slow attack; in the second of fast attack. For comparison see \cite{Verge}.}
\end{figure}

The trachea starts to oscillate independently and before the acoustical ones until to reach the stationary regime when all vocal apparatus globally self-oscillates. The time delay between the traces in the specific case of [e] vowel is equal to 0.035s for a female corresponding to a velocity of about 30m/s (reminding that the two lasers are displaced of 0.11m); and it is $4*10^{-4}s$ indicating a velocity of propagation of about $270m/s$ for a man. More in general, the time delay depends on the type of vowel, on the anatomic differences between males and females, and it is sensitive to the steepness to the rise of the supply pressure buildup due to the volume injection.

This situation reminds us the behaviour of the transient attack in a playing organ pipe. As in that case, we can speak of slow and fast attack: The form of transient attack depends upon the characteristic of the pipe and the geometry of the jet, and upon the time variation of the air pressure producing the jet (for comparison see \cite{Verge}).
You can use this analogy in order to further clarify: in a conventional theory, it is expected that the trachea would play the same role of the pipe foot, which only allows air to flow into the resonator as a thin jet of wind directed towards the mouth.
But the measurement of acoustical and mechanical vibrations (CH1) in a playing organ pipe display a different behaviour with respect to the voice. Pointing the laser beam to the foot (CH1) and simultaneously measuring the acoustic field one observes that the acoustic signal starts before CH1 (see Fig.6). This is in agreement with the physical process generating the sound in an flue organ pipe. When a jet leaves a slit in the foot and impinges against an edge, a tone is generated. This movement of air around the upper lip provides the excitation of the air column resonating inside the pipe body and induces the external acoustic field recorded by the microphone as first arrival. The vibration is then transferred to the foot (CH1) when the stationary phase displays an overall self-oscillation. The pipe foot and its length do not modify the pitch, which on the contrary depends on the length and the volume of the resonator \cite{Fletcher,Delauro}. If the trachea plays the same role of the foot, it should be noted that the onset of signal does start after the acoustic one. This indicates that the role of the trachea in the vowels production as well as in the pitch formation has to be taken into account.

 \begin{figure}
\label{simulation}
\includegraphics[width=12cm]{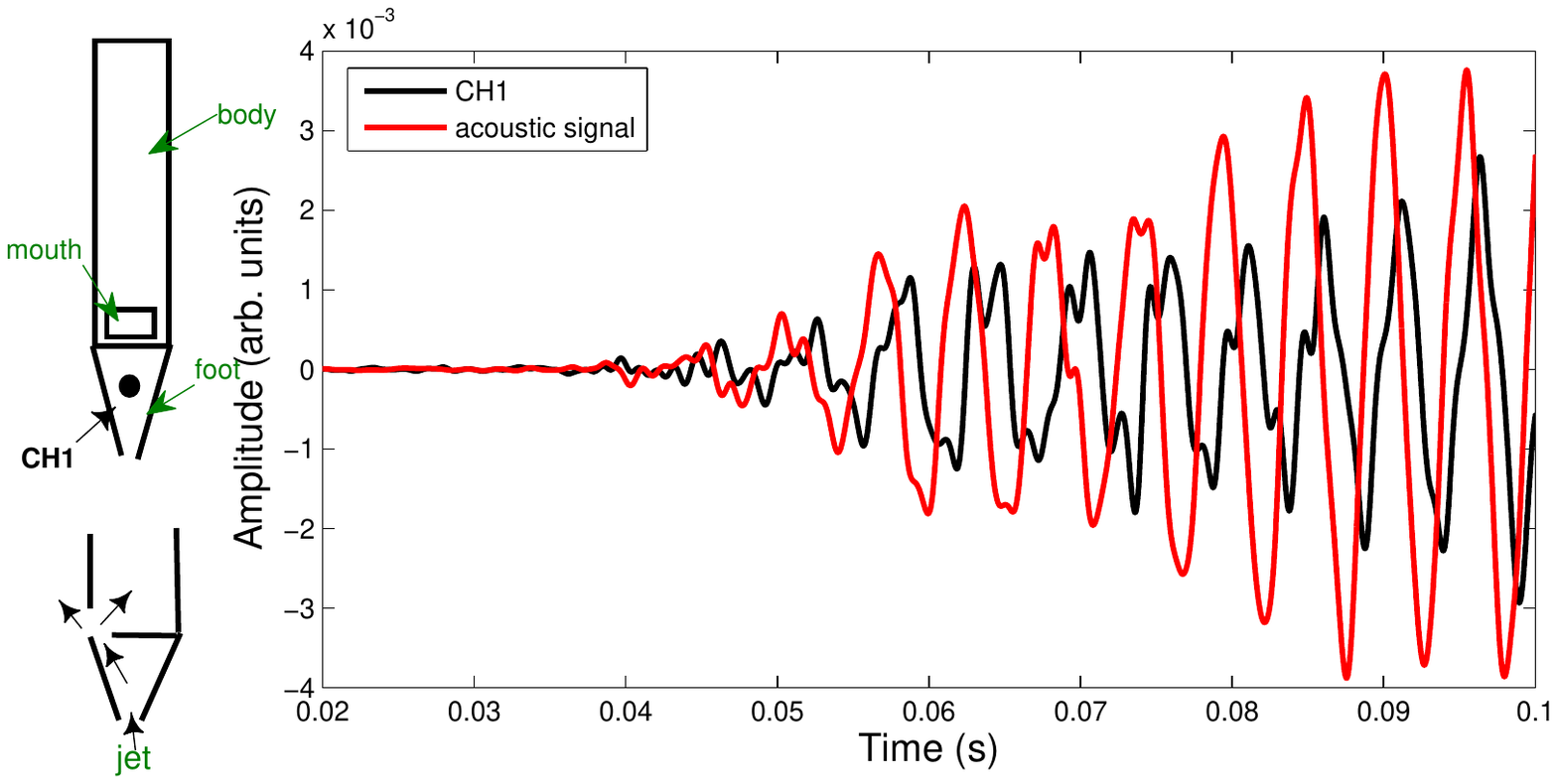}\\
\caption{Sketch of the flue organ pipe and superposition of acoustical and mechanical signals associated to a playing organ pipe. The acoustical signal starts before the mechanical one recorded pointing the laser beam to the foot (CH1).}
\end{figure}

\section{Discussions and conclusions}
We have performed several experiments to investigate the acoustical and mechanical vibrations associated with
the human vocal apparatus when producing Italian vowels. A well-known not-invasive optical measurement technique has been used, i.e., laser Doppler vibrometer. The observed oscillations are epiphenomena of many relevant processes that occur alongside or in parallel in human body (i.e., respiratory and cardiovascular systems). Focusing the attention on the frequency range of the sound production, we observe the mechanical tracheal oscillations due to a time varying quasi periodic air pulse. In literature, the latter is generally attributed to differences in pressure between sub and supra glottal tract. Comparing the acoustical and mechanical transient attacks at 2-points recorded simultaneously, we show that the trachea starts to oscillate independently and before the acoustical ones. In addition, the mechanical oscillation of the trachea starts before that of larynx indicating that there is a contribute not attributable to difference in the pressure across the vocal folds but due to an already oscillating source.
According with literature, the geometrical configurations exhibiting self-sustained oscillations are various and also include a jet issuing from a nozzle into a cavity in a sudden expansion. In the vowels production, a basic instability occurs at tracheal level.
These tracheal oscillation is the signature, in other words the pitch, that help the ear to recognize who is producing the vowel. Indeed, playing the tracheal oscillation, the speaker is clearly recognized. That signal has not an impulsive characteristic as previously hypothesized for sub/supra glottal flow but it is a quasi-periodic oscillation of trachea-lung system. We show these oscillations are generated by a deterministic low-dimensional system that represents the effective coarse-grained description associated to the fluid dynamical one. The representative dynamical system is nonlinear and low-dimensional for the trachea and represents the first part of a more complex system generating vowels.
In conclusion, the instabilities of trachea have to be added to those of glottis to take advantage in defining a more refined model of voice production.


\end{document}